\documentclass[letterpaper, 10 pt, conference]{ieeeconf}  % Comment this line out if you need a4paper

\usepackage{graphicx}
\usepackage{times}
\usepackage{amsmath,amsbsy,amssymb,mathtools,marvosym}
\usepackage{color,soul}
\usepackage[us]{datetime}
\usepackage{url}
\usepackage{theorem}
\usepackage{multirow}
\usepackage{footnote}
\usepackage{nomencl}
\usepackage{filecontents}
\usepackage{cite}
\usepackage{stackengine}
\usepackage{todonotes}
\usepackage{booktabs}

\usepackage{commath}
\usepackage{dsfont}

\theoremstyle{plain}
\newtheorem{theorem}{Theorem}

\newtheorem{remark}{Remark}

\usepackage{hyperref}

\usepackage{algorithmic}
\usepackage[ruled,vlined, linesnumbered]{algorithm2e}
\newcommand{\negpar}[1][-1em]{%
  \ifvmode\else\par\fi
  {\parindent=#1\leavevmode}\ignorespaces
}

\IEEEoverridecommandlockouts                              % This command is only needed if 
\title{
Deep Reinforcement Learning for Reach-Avoid-Stay Problems
}

\author{Gabriel Chenevert, Jingqi Li, Achyuta Kannan, Sangjae Bae, and Donggun Lee% <-this % stops a space
\thanks{Gabriel Chenevert, and Donggun Lee are with the Department of Mechanical and Aerospace Engineering, North Carolina State University, Raleigh, USA.
       {\tt\small \{gechenev,donggun\_lee\}@ncsu.edu}. 
Jingqi Li is with the Department of Aerospace Engineering and Engineering Mechanics, at University of Texas, Austin, USA.
       {\tt\small jingqi.li@austin.utexas.edu}. Achyuta Kannan is with the Department of Computer Science , North Carolina State University, Raleigh, USA.
       {\tt\small ackannan@ncsu.edu}. Sangjae Bae is with Honda Research Institute, 
       {\tt\small sbae@honda-ri.com}}
   \thanks{This research is supported by the Honda Research Institute.}
}

\begin{document}
\maketitle
\thispagestyle{empty}
\pagestyle{empty}

%%%%%%%%%%%%%%%%%%%%%%%%%%%%%%%%%%%%%%%%%%%%%%%%%%%%%%%%%%%%%%%%%%%%%%%%%%%%%%%%
\begin{abstract}
Reach-Avoid-Stay (RAS) tasks are essential in applications where systems must safely reach a target set and remain within it under all bounded disturbances. Existing approaches either struggle to compute the maximal robust RAS set—the set of all states from which the RAS task is achievable—or are limited in handling general dynamic systems. To address these challenges, this paper proposes a two-step deep reinforcement learning framework that jointly learns the maximal robust RAS set and the corresponding control policy. The first step identifies the maximal robust control-invariant set within the target set and derives a policy that ensures the system remains within it. The second step computes the maximal robust reach-avoid (RA) set using this invariant set as the target, and it is proven that this RA set is equivalent to the maximal robust RAS set. Leveraging this result, a switching policy is constructed from the two step-wise policies, which constitutes a valid policy guaranteeing completion of the RAS task. Simulation results demonstrate that the proposed framework (1) computes the exact maximal robust RAS set in the absence of training errors, yielding the least restrictive RAS policy, and (2) identifies the RAS set with high accuracy while outperforming baseline methods on RAS tasks.
\end{abstract}

\section{Introduction}
\label{sec:intro}

% General applications + need for reachability
Modern robotic motion planning and control methods aim to generate safe trajectories in complex environments. A prominent approach is reachability analysis \cite{bansal2017hamilton}, which determines the set of states from which a control policy exists to safely accomplish a given task. Reachability analysis has been applied to diverse systems, including aircraft collision avoidance \cite{prandini_application_2008}, human-robot interaction \cite{bansal2020hamilton}, and robot manipulation \cite{jauhri2022robot}. Among reachability formulations, the \emph{reach-avoid} (RA) problem is one of the most widely used: the goal is to drive a system to a target while satisfying safety constraints. Recent advances in RA methods involve leveraging deep learning \cite{bansal2021deepreach} and reinforcement learning (RL) \cite{allen_machine_2014, dawson_safe_2021, hsu_safety_2021, hsu_safety_2023, hsu_sim--lab--real_2022, hsu_isaacs_2022, fisac_bridging_2019, li_certifiable_2024} to enhance computational efficiency for high-dimensional systems. Other work has extended RA to stochastic processes \cite{summers_stochastic_2011}.

% RA can not stay
However, RA methods can cause the system to leave the target set \cite{li_infinite-horizon_2024}. This occurs when the target strictly contains and is not equivalent to the robust viability kernel, the maximal robust control-invariant subset of the target set under all bounded disturbances. If the system reaches the target set but not the robust viability kernel, it will exit the target set. In vehicle applications, for example, this tends to occur when the system enters the target set with excessive velocity and lacks sufficient control to slow and remain within it. RA methods \cite{li_infinite-horizon_2024} are unable to determine which states are in the robust viability kernel, and are therefore insufficient for tasks where staying within the target set is critical.

% Problem statement
Reachability tasks that require staying within the target set have motivated the introduction of the \emph{reach-avoid-stay} (RAS) problem. The RAS problem seeks the set of initial states and a corresponding policy such that the system can be driven safely to the target set and kept there indefinitely despite all bounded disturbances. This set is called the \emph{maximal robust RAS set}. Together with its associated policy, the maximal robust RAS set provides a guarantee that the RAS task can be achieved from the \textit{largest} set of states. Without this set, the policy may exclude feasible states or mistakenly include unsafe ones. False positives, states incorrectly classified as belonging to the RAS set, are dangerous, as the task can not be safely completed from those states, leading to safety violations. False negatives exclude feasible states, leading to unnecessarily conservative policies.

\begin{figure}[t]
    \centering
    \includegraphics[width=\linewidth]{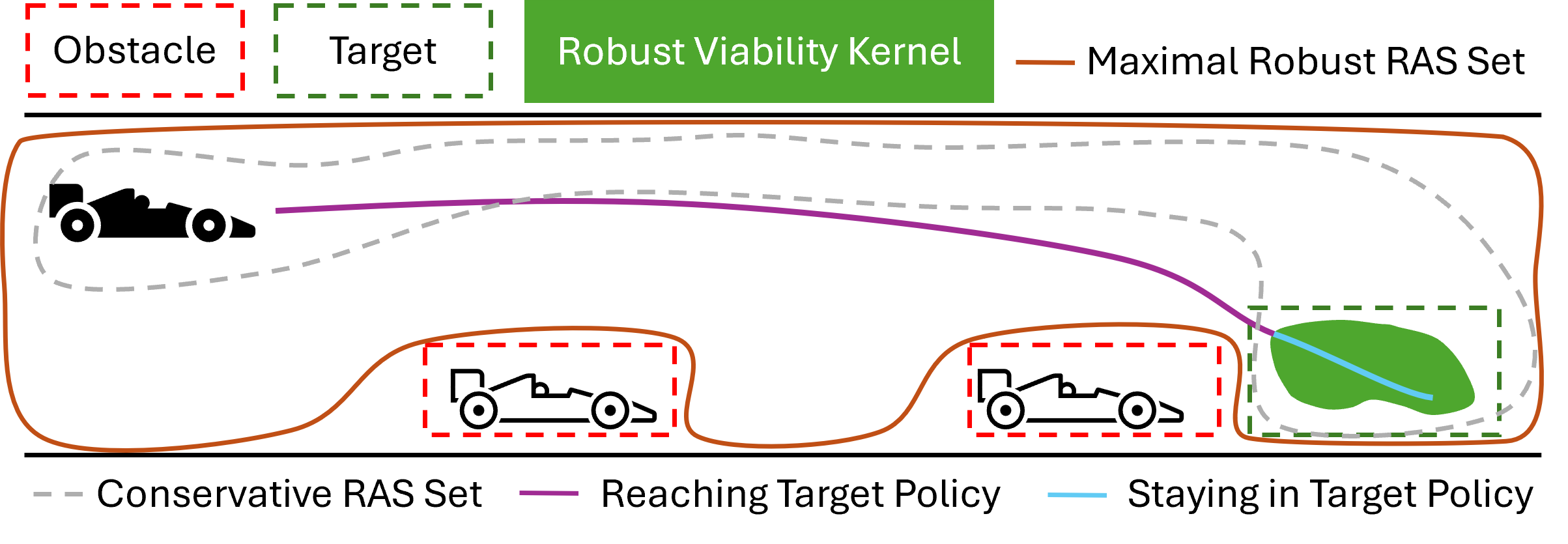}
    \caption{The reach-avoid-stay (RAS) task involves guiding a system from an initial state, to a target set where it remains while safely avoiding all obstacles against bounded disturbances. The maximal robust RAS set (orange region) is the set of all states where the RAS task can be performed. The robust viability kernel, the solid green region, is the set of all states which can safely stay within the target. A RAS policy must first direct the system from within the RAS set to the robust viability kernel, and then remain within the robust viability kernel. If a RAS method has only identified the conservative RAS set (gray dashed line), it may encounter false negative safety classifications, restricting the range of the control policies.
    }
    \label{fig:rasConcept}
    \vspace{-2em}
\end{figure}

% Previous RAS Methods
Several methods have been developed for the RAS problem; however, each approach encounters distinct challenges. Control-Lyapunov-Barrier function (CLBF)-based methods \cite{meng_lyapunov-barrier_2022} require the design of CLBFs, which is challenging for high-dimensional or nonlinear systems. Reinforcement learning-based approaches \cite{dawson_safe_2021, so_how_2024} attempt to address the challenges of designing CLBFs, but fail to identify the maximal robust RAS set, resulting in inaccurate safety classification and conservative policies. Funnel-based control methods \cite{das_funnel-based_2023, das_prescribed-time_2024} circumvent the challenging CLBF design process, but assume that the system can stop instantly. These limitations cause current methods for the RAS problem to be either difficult to design, unable to identify the maximal RAS set, or dependent on restrictive assumptions about dynamical systems that limit their general applicability.

% Our Approach 
To address the above limitations, we propose a two-step deep-RL framework for RAS problems that computes the maximal robust RAS set and its corresponding policy. First, we compute the robust viability kernel within the target set. Second, we apply the RA method \cite{li_infinite-horizon_2024}, replacing the target set with the robust viability kernel. We introduce a value function transformation that yields a function strictly positive within the robust viability kernel and non-positive elsewhere, enabling it to serve as a reward function for the RA method. Finally, we design a switching policy for RAS tasks by combining the policies from both steps.

% Contribution
The primary contributions of our learning framework are:
\begin{enumerate}

    \item \textbf{Theoretical Contribution:} We prove that our RAS framework identifies the maximal robust RAS set in the absence of training errors, and that the resulting RAS policy achieves the RAS task for all states in this set under all bounded disturbances. These results are enabled by our proposed value function transformation, without which the integration of the two prior reachability frameworks cannot solve the RAS problem.
    
    \item \textbf{Learning an Accurate Maximal Robust RAS Set Via Deep-RL}: We develop a deep-RL framework to learn the maximal robust RAS set for high-dimensional systems, and demonstrate that even with the training error introduced by deep-RL methods, our framework characterizes the maximal robust RAS set with high accuracy.
    \item \textbf{Practical Applicability}: Our approach handles complex environments for which designing CLBFs \cite{meng_lyapunov-barrier_2022} is difficult, and does not rely on the instant stopping assumption made by funnel-based methods \cite{das_funnel-based_2023, das_prescribed-time_2024}, enabling broader applicability.
    \item \textbf{Empirical Validation}: We conduct simulations to validate our framework’s effectiveness. Our approach outperforms CLBF-based methods \cite{meng_lyapunov-barrier_2022} at identifying the maximal robust RAS set, reducing false negatives, yielding a less cautious policy while retaining safety. It also achieves higher success rates for the RAS task compared to a reach-avoid method \cite{li_certifiable_2024}.
\end{enumerate}

\section{Background: Reach-Avoid Analysis}
We consider nonlinear dynamical systems, described by 
\begin{equation}
    \mathrm{x}_{t+1} = f(\mathrm{x}_t,u_t,d_t),
    \label{eq:dynamics}
\end{equation}
where $\mathrm{x}_t\in\mathbb{R}^n$ is the state, $u_t\in U \subset \mathbb{R}^{m_u}$ is control, and $d_t\in D\subset\mathbb{R}^{m_d}$ is a disturbance at time $t$. For robust analysis, we consider an adversarial disturbance \cite{hsu_isaacs_2022}, which can model environmental uncertainty and bounded discrepancies between the simulation and the physical system. If our control policy succeeds under such conditions, it is guaranteed to succeed under less severe disturbances.

Consider an open set $T\subseteq\mathbb{R}^{n}$ representing the target set. We define a Lipschitz continuous bounded reward function $g$ which indicates if a state within the target set:
\begin{equation}
    \label{eq:gx}
    g(x) > 0 \Leftrightarrow x \in T.
\end{equation}
Similarly, we consider an open set $C\subseteq\mathbb{R}^{n}$ representing the constraint set (safe region). We define an additional Lipschitz continuous bounded reward function exists such that:
\begin{equation}
    \label{eq:lx}
    l(x) > 0 \Leftrightarrow x \in C.
\end{equation}

We present the RA problem from prior work in \cite{li_certifiable_2024} and \cite{li_infinite-horizon_2024}. The objective is to identify the maximal robust RA set along with its corresponding control policy $\pi_u:\mathbb{R}^n\rightarrow U$  such that for all disturbance policies, $\pi_d:\mathbb{R}^n\rightarrow D$, the system reaches the target set while remaining within the constraint set. The resulting maximal robust RA set is defined as:
\begin{align}
    \hspace{-0.2em}\mathcal{RA} \coloneqq  \left\{ x \bigg| 
     \begin{tabular}{l}
     $\exists \pi_u$ s.t. $\forall \pi_d, \exists \tau\in\{0,...\}$ s.t. \\
        1)  $\mathrm{x}_t\in C, t=\{0,...,\tau\}$ and \\
        2)  $\mathrm{x}_\tau\in T$
    \end{tabular}\right\},
\label{eq:RA_Set_Definition}
\end{align}
where $\mathrm{x}_{t+1}= f(\mathrm{x}_t, \pi_u (\mathrm{x}_t), \pi_d(\mathrm{x}_t ) ), \text{ and } \mathrm{x}_0 =x.$. While the RA task only requires that the trajectory reaches $T$ momentarily, the RAS task requires it to remain in $T$ for all time after the reaching time $\tau$. Therefore, we extend the RA problem to RAS, which is the core problem of our work.

\section{Problem Definition: Reach-Avoid-Stay}
\label{sec:RAS}
Consider the dynamical system described in \eqref{eq:dynamics}, along with the target set $T$ defined in \eqref{eq:gx}, and the constraint set $C$ defined in \eqref{eq:lx}. The RAS problem is to identify the maximal robust RAS set and corresponding control policy, $\pi_u:\mathbb{R}^n\rightarrow U$ such that for all disturbance policies, $\pi_d:\mathbb{R}^n\rightarrow D$, the system remains within the constraint set and stays in the target set after reaching it. The maximal robust RAS set is defined as:

\begin{align}
    \hspace{-0.2em}\mathcal{RAS} \coloneqq  \left\{ x \bigg| 
     \begin{tabular}{l}
     $\exists \pi_u$ s.t. $\forall \pi_d,$ \\
        1)  $\mathrm{x}_t\in C, t=\{0,...\}$ and \\
        %\quad\quad 
        2) $\exists \tau\in\{0,...\}$   s.t. $\mathrm{x}_t\in T, t=\{\tau,...\}$
    \end{tabular}\right\},
\label{eq:RAS_Set_Definition}
\end{align}
\normalsize    
where $\mathrm{x}_t$ is governed by a RAS policy $\pi_u$ and a disturbance policy $\pi_d$:
\begin{align}
    \mathrm{x}_{t+1}= f(\mathrm{x}_t, \pi_u (\mathrm{x}_t), \pi_d(\mathrm{x}_t ) ), \text{ and } \mathrm{x}_0 =x.
    \label{eq:dynamics_policy}
\end{align}
The first condition in \eqref{eq:RAS_Set_Definition} encodes the safety condition, while the second encodes the reach and stay conditions.

\section{Two-Step Deep Deterministic Policy Gradient for RAS}
\label{sec:optCtrl}

In this section, we propose a two-step RL framework to compute $\mathcal{RAS}$ and the corresponding policy. Our approach integrates the RL method for RA tasks \cite{li_certifiable_2024} and the robust viability kernel \cite{aubin_regulation_2011}. In the first step, we identify the robust viability kernel set within the target set and a control policy for remaining within it. In the second step, we apply the RA formulation, but replace the target set with the robust viability kernel specified in the first step and identify a policy for reaching the robust viability kernel. We prove that the resulting $\mathcal{RA}$ is equivalent to $\mathcal{RAS}$ and that this $\mathcal{RAS}$ is maximal and robust; no state outside $\mathcal{RAS}$ subject to adversarial disturbances can achieve the RAS task. Section \ref{subsec:h} presents the robust viability kernel analysis. Building on this result, in Section \ref{subsec:v}, we propose our main framework: the RAS formulation, along with its proof and algorithms.

\subsection{Computing the Robust Viability Kernel}
\label{subsec:h}
In this subsection, we leverage the discounted formulation from \cite{li_certifiable_2024} to find the largest robust control-invariant set within the target set while avoiding obstacles. This set is referred to as the robust viability kernel:
\begin{align}
    \mathcal{AS} \coloneqq \left\{ x~\bigg|~
    \begin{tabular}{l}
    $\exists \pi_u \textrm{ s.t. } \forall \pi_d,$\\ 
    $\mathrm{x}_t\in C\cap T , t=\{0,...\}$
    \end{tabular}\right\}.
    \label{eq:AS_Set_Definition}
\end{align}

We define a discounted value function, $H$, that characterizes the robust viability kernel within the target set while remaining in the constraint set:
\begin{align}
    H(x) \coloneqq \sup_{\pi_u}\inf_{\pi_d} \inf_{t=\{0,...\}}\gamma^t \bar{g}(\mathrm{x}_t),
    \label{eq:Def_H}
\end{align}
where $\bar{g}$ characterizes the target set excluding the avoid region:
\begin{align}
    \bar g (x)\coloneqq \min\left( g(x), l(x ) \right),
\end{align}
and $\gamma\in(0,1)$ is the discount factor.
The value of $\bar g$ is positive if and only if the state is within both the target set and the constraint set. $H$ evaluates the discounted value of $\bar g$ across an infinite time horizon. If at any state is outside the target set or the constraint set, $H$ will be negative, indicating that the initial state was outside of $\mathcal{AS}$.

The following remark presents the Bellman equation for $H$ that characterizes $\mathcal{AS}$ and the corresponding policy that enables the system to safely stay in the target if the state is in $\mathcal{AS}$.

\begin{remark}
\label{remark:H}
    The discounted-value function $H$, defined in \eqref{eq:Def_H}, is the unique solution to the following Bellman equation \cite{lewis_optimal_2012}:
    \begin{align}
        H(x) = \min\left\{ \bar{g}(x), \gamma \max_{u\in U}\min_{d\in D} H(f(x,u,d))  \right\},
        \label{eq:hbellman}
    \end{align}
    since the corresponding Bellman operator is a contraction mapping \cite{denardo_contraction_1967}.
    The robust viability kernel set within the target set while avoiding unsafe regions is the level-zero set of $H$:
    \begin{align}
        \mathcal{AS} = \{ x ~|~ H(x)=0\}.
    \end{align}
    Note that $H$ is non-positive for all $x\in\mathbb{R}^n$: $H(x)=0$ for $x\in\mathcal{AS}$ and $H(x)<0$ for $x\notin\mathcal{AS}$.
    Also, for $x\in \mathcal{AS}$, any control that satisfies the following equation enables staying in the target set and constraint set for any bounded disturbance:
    \begin{align}
        \pi_{H}(x) \in \arg\max_{u\in U} \min_{d\in D} H( f(x,u,d)),
    \end{align}
    where $\pi_H:\mathbb{R}^n\rightarrow U$ is a policy that keeps trajectories safe indefinitely if they start in $\mathcal{AS}$.    
\end{remark}

\subsection{Computing the Maximal Robust RAS Set}
\label{subsec:v}
In this subsection, we utilize the RA formulation from \cite{li_certifiable_2024} to specify $\mathcal{RAS}$ and its corresponding policy by replacing $T$ in \eqref{eq:RA_Set_Definition} with $\mathcal{AS}$. For this process, we need to use a value function whose super-zero-level set is $\mathcal{AS}$ to provide an incentive to reach $\mathcal{AS}$ as quickly as possible. Without this, in RA analysis the system receives the same reward whether it reaches $\mathcal{AS}$ in finite time or only when time is very large approaching infinity. In such a case, the policy does not actively drive the system toward $\mathcal{AS}$. Therefore, since $H$ is identically zero on $\mathcal{AS}$, it is not a suitable candidate.

To provide a reaching incentive, we design a new value function $H_g: \mathbb{R}^n \rightarrow \mathbb{R}$ as follows.
\begin{align}
    H_g (x) \coloneqq \begin{cases}
        H(x)~\text{in}~\eqref{eq:Def_H}, & \text{if } H(x)<0, \\ g(x)~\text{in}~\eqref{eq:gx}, & \text{otherwise.}
    \end{cases}
    \label{eq:Def_Hg}
\end{align}
Theorem \ref{theorem:positiveHg} shows that the super-zero-level set of $H_g$ represents the same robust viability kernel $\mathcal{AS}$, while the values inside the set are strictly positive.
\begin{figure}
    \centering
    \includegraphics[width=1\linewidth]{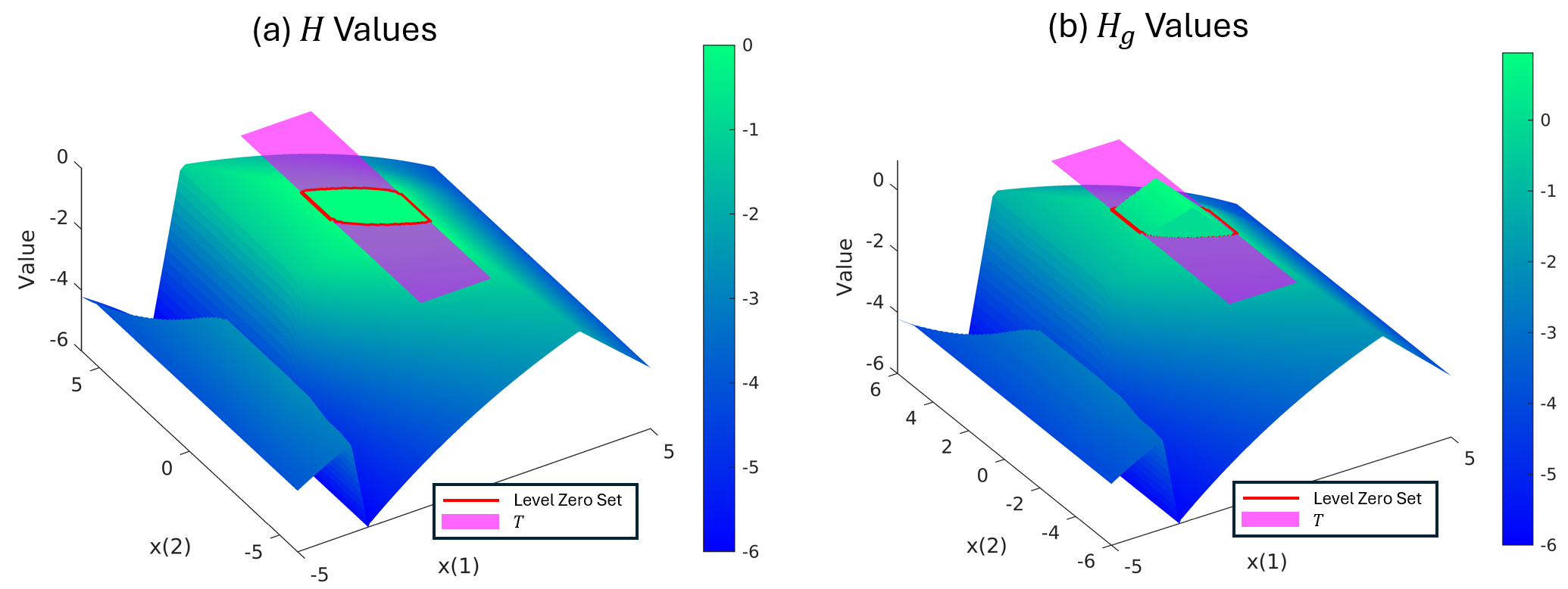}
    \caption{ These figures show the transformation from $H$ shown in (a) to $H_g$ shown in (b) for the cart example whose dynamics and target set formulation are available in Section \ref{subsec:cart}. The zero-level set of $H$ is the same as the super-zero-level set of $H_g$. $H_g$ is strictly positive in $\mathcal{AS}$ as demonstrated in Theorem \ref{theorem:positiveHg}.}
    \label{fig:hgx}
\vspace{-2em}
\end{figure}

\begin{theorem}
$H_g$ is positive if and only if $x$ is within $\mathcal{AS}$.
\label{theorem:positiveHg}
\end{theorem}    
\begin{proof}
Consider two cases: $x\in\mathcal{AS}$ and $x\notin\mathcal{AS}$.

\noindent Case 1: $x\in\mathcal{AS}$. Since Remark \ref{remark:H} implies $H(x)=0$, we have $H_g(x)=g(x)$ by the definition of $H_g$ in \eqref{eq:Def_Hg}. By the definition of $\mathcal{AS}$ in \eqref{eq:AS_Set_Definition}, $g(x)>0$. Thus, $H_g(x)=g(x)>0$.

\noindent Case 2: $x\notin\mathcal{AS}$. By Remark \ref{remark:H} and \eqref{eq:Def_H}, it follows that $H(x)<0$. Then, by \eqref{eq:Def_Hg}, we conclude that $H_g(x)=H(x)<0$.
\end{proof}
An example of $H_g$ constructed for the cart example in Section \ref{subsec:cart} can be found in Fig. \ref{fig:hgx} (b).

To characterize $\mathcal{RAS}$, we use $H_g$ to define another discounted-value function $V$:

\small{
\begin{multline}
\label{eq:vValue}
V(x)\coloneqq \sup_{\pi_u} \inf_{\pi_d} \sup_{t=\{0,...\}} \min\{ \gamma^t H_g(\mathrm{x}_t),\inf_{s=\{0,...,t\}} \gamma^s l(\mathrm{x}_s) \}.
\end{multline}
}
\normalsize

\noindent Subsequently, the Bellman equation for $V$ and the corresponding policy can be derived as described by the following remark.
\begin{remark}
    The value function $V$, defined in \eqref{eq:vValue}, is the unique solution to the following Bellman equation \cite{li_certifiable_2024}:
    \begin{align}
        V(x) = \min\{l(x), \max\{ H_g(x), \gamma \max_{u\in U} \min_{d\in D} V(f(x,u,d)) \}\},
    \end{align}
    since the corresponding Bellman operator is a contraction mapping \cite{li_certifiable_2024}.
    Any policy that satisfies the following equation enables reaching the robust viability kernel within the target set $\mathcal{AS}$ while satisfying the constraint set:
    \begin{align}
        \pi_V(x) \in \arg\max_{u\in U} \min_{d\in D} V( f(x,u,d)),
    \end{align}
    where $\pi_V:\mathbb{R}^n\rightarrow U$ is the corresponding RA policy \cite{li_certifiable_2024}.
\end{remark}\vspace{-1em}

Now, we present our main theorem, which states that the value function $V$ characterizes the maximal robust RAS set $\mathcal{RAS}$~\eqref{eq:RAS_Set_Definition}.
\begin{theorem}
The super-zero-level set of $V$ \eqref{eq:vValue} is the maximal robust RAS set, $\mathcal{RAS}$ \eqref{eq:RAS_Set_Definition}.
    \begin{align}
        \mathcal{RAS} = \{ x ~|~ V(x) > 0\}
    \end{align}
    
    \label{theorem:mainThm}
\end{theorem}
\begin{proof}
    (i) Suppose $x\in\mathcal{RAS}$. Then there exists $\pi_{u}^*$ such that, for all $\pi_d$, $l(\mathrm{x}^*_t)>0$, $t=0,...,$ and there exists $\tau^*$ such that, $g(\mathrm{x}^*_t)>0$, $t=\tau,...$, where $\mathrm{x}^*$ solves \eqref{eq:dynamics_policy} for $\pi_u^*$ and $\pi_d$. Note that $\tau^*$ and $\mathrm{x}^*$ depend on $\pi_d$. These conditions can be rewritten as follows: for all $\pi_d$, there exists $\tau^*$ such that $l(\mathrm{x}^*_t)>0$ for $t=0,...,\tau^*$ and $\bar g(\mathrm{x}^*_t)>0$ for $t=\tau^*,...$, which implies that $\mathrm{x}^*_{\tau^*}\in \mathcal{AS}$. By \eqref{eq:Def_Hg}, $H_g(\mathrm{x}^*_{\tau^*})=g(\mathrm{x}^*_{\tau^*})>0$.
    Therefore, we have
    \begin{align}
        \inf_{\pi_d} \sup_{\tau=\{0,...\}} \min\{ \gamma^\tau H_g(\mathrm{x}^*_{\tau}), \min_{t=\{0,...,\tau\}} \gamma^t l(\mathrm{x}^*_t)  \} >0.
        \label{eq:thmRAS_proof_eq1}
    \end{align}    
    Since $\pi_u^*$ may not be a maximizer for $V$,
    \begin{align}
        V(x)& \geq \inf_{\pi_d} \sup_{\tau=\{0,...\}} \min\{ \gamma^\tau H_g(\mathrm{x}^*_\tau), \inf_{s=\{0,...,\tau\}} \gamma^s l(\mathrm{x}^*_s) \} >0.
        \label{eq:thmRAS_proof_eq2}
    \end{align}
    \\
    (ii) Suppose $V(x)>0$. Then, there exists $\pi_{V}$ such that, for any $\pi_d$, there exists $\tau^*$ such that 
    \begin{align}
        H_g(\mathrm{x}^*_{\tau^*})>0, l(\mathrm{x}^*_t)>0, t=\{0,...,\tau^*\}, 
        \label{eq:thmRAS_proof_eq4}        
    \end{align}
    where $\mathrm{x}^*$ solves \eqref{eq:dynamics_policy} for $\pi_V, \pi_d$. Since $H_g(\mathrm{x}^*_{\tau^*})$ is positive, there exists another policy $\pi_{H}$ such that, for any $\pi_d$, 
    \begin{align}
        g(\mathrm{x}_t^{H*})>0, l(\mathrm{x}_t^{H*})>0, ~t=\{\tau^*,...\},
        \label{eq:thmRAS_proof_eq3}
    \end{align} 
    where $\mathrm{x}_t^{H*}$ solves \eqref{eq:dynamics_policy} for $\pi_{H}, \pi_d$, and $\mathrm{x}_{\tau^*}^H=\mathrm{x}_{\tau^*}$. 
  Combining $\pi_V$ and $\pi_H$ into a unified policy $\pi_{\mathcal{RAS}}$ yields:
    \begin{align}
    \label{eq:switch}
        \pi_{\mathcal{RAS}}(x) = 
            \begin{cases}
                \pi_{V}(x) &\text{if } H_g(x)\leq 0 \\
                \pi_{H}(x) & \text{otherwise.}
            \end{cases}
    \end{align}
    Using $\pi_\mathcal{RAS}$, for each $\pi_d$, the corresponding state trajectory $\bar{\mathrm{x}}$ solving \eqref{eq:dynamics_policy} is 
        \begin{align}
            \bar{\mathrm{x}}_t = \begin{cases}
                \mathrm{x}^*_t & t=\{0,...,\tau^*\} \\
                \mathrm{x}^{H*}_t & t=\{\tau^*,...\}.
                \end{cases}
    \end{align}
    Therefore, $\pi_\mathcal{RAS}$ satisfies, for any $\pi_d$, $l(\bar{\mathrm{x}}_t)=l(\mathrm{x}_t)>0$ for $t\leq \tau^*$ by \eqref{eq:thmRAS_proof_eq4}, $l(\bar{\mathrm{x}}_t)=l(\mathrm{x}^{H*}_t)>0$ for $t >\tau^*$ by \eqref{eq:thmRAS_proof_eq3}, and $g(\bar{\mathrm{x}}_t)=g(\mathrm{x}^{H*}_t)>0$ for $t\geq \tau^*$ by \eqref{eq:thmRAS_proof_eq3}. Thus, $x\in \mathcal{RAS}$, and $V(x)>0$.
\end{proof}
The proof of Theorem \ref{theorem:mainThm} implies that if the state is within $\mathcal{RAS}$, the system can accomplish the RAS task under the switching policy \eqref{eq:switch}; otherwise, the task cannot be achieved.

\begin{remark}
\label{rmk:switch}

Within $\mathcal{RAS}$, $\pi_V$ drives the system to $\mathcal{AS}$, but it cannot guarantee safety once inside. In contrast, $\pi_H$ guarantees safe invariance in $\mathcal{AS}$, which necessitates switching from $\pi_V$ to $\pi_H$. The switch occurs when $H_g(x)>0$, which by Theorem \ref{theorem:mainThm} indicates that the system is in $\mathcal{AS}$. If $x \notin \mathcal{RAS}$, $\pi_V$ will be unable to safely reach $\mathcal{AS}$, thus the system will be unable to perform the RAS task.
\end{remark}

\subsection{A Two-step Deep-RL Framework for Learning the Maximal Robust RAS Set and the Associated Policies}
We summarize our two-step framework in Algorithm~\ref{alg:TWF}. First, we learn a value function $H$ that characterizes $\mathcal{AS}$, along with the corresponding stay policy $\pi_{H}$. Second, to encourage the system to reach the target set, we construct $H_g$ according to \eqref{eq:Def_Hg}, which assigns positive values within $\mathcal{AS}$ and negative values outside it. Third, we learn a value function $V$ that characterizes the RAS set and derive the associated reach-avoid policy $\pi_V$.

To address general high-dimensional nonlinear systems, we implement Algorithm~\ref{alg:TWF} using Deep Deterministic Policy Gradient (DDPG) \cite{li_robust_2019,lillicrap_continuous_2019,silver_deterministic_2014}. For low-dimensional systems, where the curse of dimensionality has less impact, DDPG can be replaced with tabular Q-learning \cite{sutton_reinforcement_2020} or similar optimization methods.Guidance on training specific to our examples is provided in Section~\ref{sec:twoStepDDPG}.

\begin{savenotes}
\begin{algorithm}[t!]
    \SetAlgoLined
    \caption{A Two-step Framework for RAS Problems:}
    \label{alg:TWF}
    \tcp{This training can be performed using DDPG or any value iteration based method.}
    
    \textbf{Train an approximation function $H^\phi$, a control policy $\pi^\theta_{H}$, and a disturbance policy $\pi^\rho_{d_H}$:}
    Update $H^\phi$ by minimizing the Bellman error:
    \begin{equation}
        \|H^\phi(x) - \min\{ \bar{g}(x), \gamma H^\phi(f(x,\pi^\theta_{H}(x),\pi^\rho_{d_H}(x)))  \} \|_2.
    \end{equation}
    Use $H^\phi\Big(f\big(x,\pi^\theta_{H}(x),\pi^\rho_{d_H}(x)\big)\Big)$ to train $\pi^\theta_{H}$ and $\pi^\rho_{d_H}$ such that $\pi^\theta_{H}$ maximizes $H^\phi$ and $\pi^\rho_{d_H}$ minimizes $H^\phi$.
    
    \textbf{Create $H^\phi_g$:} Combine $H^\phi$ with $g$ according to~\eqref{eq:Def_Hg}.\\

    \textbf{Train an approximation function $V^\varphi$, a control policy for $\pi^\vartheta_{V}$, and a disturbance policy $\pi^\varrho_{d_V}$:}
    Update $V^\varphi$ by  minimizing the Bellman error:
    \begin{multline}
        \| V^{\varphi}(x) - \min\{l(x), \max\{ H_g^\phi(x),
        \\ \gamma V^\varphi(f(x,\pi^\vartheta_{V}(x),\pi^\varrho_{d_V}(x))) \}\}\|_2.
    \end{multline}
    Use $V^\varphi\Big(f\big(x,\pi^\vartheta_{V}(x),\pi^\varrho_{d_V}(x)\big)\Big)$ to train $\pi^\vartheta_{V}$ and $\pi^\varrho_{d_V}$ such that $\pi^\vartheta_{V}$ maximizes $V^\varphi$ and $\pi^\varrho_{d_V}$ minimizes $V^\varphi$.
\end{algorithm}
\end{savenotes}

\section{Simulation}
\label{sec:examples}
In this section, we demonstrate the following:

\begin{itemize}
    \item \textbf{Claim 1:} Our framework identifies the maximal robust RAS set, $\mathcal{RAS}$,  if there is no training error. We verify that this is a super-set of the RAS set identified by the CLBF baseline \cite{meng_lyapunov-barrier_2022} resulting in more accurate safety classification and less conservative policies. (Section \ref{subsec:cart})
    
    \item \textbf{Claim 2:} Even with training error, our framework identifies $\mathcal{RAS}$ with over 95\% accuracy. (Section \ref{subsec:baseline})

    \item \textbf{Claim 3:} Our framework achieves a higher success rate than a baseline RA method \cite{li_infinite-horizon_2024} in high-dimensional systems. (Section \ref{subsec:baseline})
\end{itemize}

For Claim 1, to remove training error induced by neural networks, we consider a low-dimensional system that is tractable for our framework using tabular Q-learning \cite{watkins_q-learning_1992}. We construct a CLBF \cite{meng_lyapunov-barrier_2022} to identify a RAS set for comparison in Section \ref{subsec:cart}.

To validate Claims 2 and 3, we present two high-dimensional systems, the descriptions of which are provided in Section \ref{subsec:complex}. The analysis of our framework's performance in comparison to a baseline RA method \cite{li_infinite-horizon_2024} is presented in Section \ref{subsec:baseline}. We are unable to use a CLBF-based or funnel-based method as baselines for the high-dimensional systems since designing a CLBF found in \cite{meng_lyapunov-barrier_2022} would be impractical, and they violate the funnel method assumption that the system can instantly stop \cite{das_funnel-based_2023}.

\subsection{Two-Dimensional Example: Double Integrator}
\label{subsec:cart}
This subsection demonstrates that our framework provides the maximal robust RAS set compared to another set computed by a CLBF baseline method \cite{slotine1991applied}.

Consider a two-dimensional cart system: 
\begin{align}
    & \mathrm{x}_{t+1} (1) = \mathrm{x}_{t} (1) + \Delta t \mathrm{x}_t(2) + \Delta t^2 (u_t + d_t), \\
    & \mathrm{x}_{t+1} (2) = \mathrm{x}_{t} (2) + \Delta t (u_t+d_t),
\end{align}
where $\mathrm{x}_t(1)$ is the position of the cart along a track at time $t$, $\mathrm{x}_t(2)$ is the velocity of the cart at time $t$, and both control ($u_t\in[-3,3]$) and disturbance ($d_t\in[-2,2]$) are acceleration. A target is located at $x(1)=0$, and an obstacle is located at $x(1)=-3$. Both the obstacle and the target have a radius of one. We design one reward function $g$ and one constraint function $l$ as follows:
\begin{equation}
    % \label{eq:cartG}
    g(\mathrm{x}_t) = 1 - |\mathrm{x}_t(1) - 0|,\quad l(\mathrm{x}_t) = |\mathrm{x}_t(1) - 3| - 1.
\end{equation}

\begin{figure}
    \centering
    \includegraphics[width=.8\linewidth]{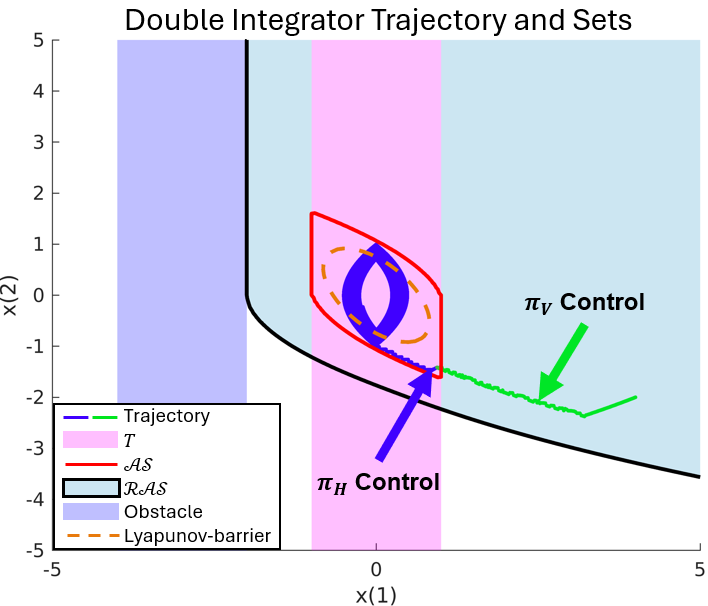}
    \caption{This figure illustrates $\mathcal{RAS}$ (within the black lines) and a trajectory driven by $\pi_\mathcal{RAS}$ in green and blue. The trajectory successfully achieves the RAS task under adversarial disturbances. $\pi_V$ is used to reach $\mathcal{AS}$ where control is switched to $\pi_H$ to enable staying.}
    \label{fig:cmd}
    \vspace{-1.5em}
\end{figure}

\textbf{Demonstration of Claim 1:} Our framework provides $\mathcal{RAS}$, a superset of the RAS set provided by the CLBF in  \cite{meng_lyapunov-barrier_2022} as supported by Theorem \ref{theorem:mainThm}. The volume of $\mathcal{RAS}$ shown in Fig. \ref{fig:cmd} is 28.67, compared to the CLBF baseline of 1.98.

\textbf{Demonstration of Switching Control Policy $\pi_\mathcal{RAS}$:} $\pi_\mathcal{RAS}$ switching control policy achieves the RAS task despite being opposed by an adversarial disturbance as can be observed from the trajectory shown in Fig. \ref{fig:cmd}. The trajectory starts from within $\mathcal{RAS}$, and is driven by $\pi_V$ safely to $\mathcal{AS}$. Upon reaching $\mathcal{AS}$, $H_g$ becomes positive, and the control policy is changed to $\pi_H$ which successfully keeps the system within $\mathcal{AS}$. 

\subsection{Complex System Experiments}
\label{subsec:complex}

Here we present two high-dimensional systems to demonstrate Claim 2 that our framework identifies $\mathcal{RAS}$ with over 95\% accuracy, and Claim 3, our framework achieves a higher success rate than a baseline RA method \cite{li_infinite-horizon_2024}. 

\subsubsection{VTOL Taxi (VTOL)} 
\label{subsubsec:VTOL}
Consider a VTOL taxi in a city. The taxi must avoid collisions with buildings and other aircraft while moving to a drop-off location. Throughout this process, wind disrupts the taxi's motion. In this demonstration, there is one ego VTOL taxi and two other VTOLs which the ego VTOL must avoid.

We model this system as an 18-dimensional state \(\mathrm{x}_t = [\mathrm{p}^e_t; \mathrm{v}^e_t; \mathrm{p}^1_t; \mathrm{v}^1_t; \mathrm{p}^2_t; \mathrm{v}^2_t]\), where \(\mathrm{p}^e_t \in \mathbb{R}^3\) and \(\mathrm{v}^e_t \in \mathbb{R}^3\) represent the ego VTOL's position and velocity, respectively, at time $t$. \(\mathrm{p}^1_t\in \mathbb{R}^3\) and \(\mathrm{v}^1_t\in \mathbb{R}^3\) denote the position and velocity of the first autonomous VTOL at time $t$, and \(\mathrm{p}^2_t\in \mathbb{R}^3\) and \(\mathrm{v}^2_t\in \mathbb{R}^3\) represent those of the second autonomous VTOL. The ego VTOL uses the following dynamics.
\begin{align}
    & \mathrm{p}^e_{t+1} = \mathrm{p}^e_{t} +\Delta t  \mathrm{v}^e_{t},\\
    & \mathrm{v}^e_{t+1} = {v}^e_{t} + \Delta t  (u_t + d_t)
    \label{eq:city_dyn},
\end{align}
where $u_t\in\mathbb{R}^3$ is a velocity input to the VTOL, and the disturbance $d_t\in\mathbb{R}^2$ represents wind and dynamics modeling errors. Each of the two obstacle VTOLs is modeled with double integrator dynamics, where the control input is designed as PD control to follow a fixed target, with the acceleration bounded between -0.5 and 0.5. We use the following $l$ and $g$ equations:

\begin{align*}
    &l(\mathrm{x}_t) = \min\bigg(\|\mathrm{p}_t^e - \mathrm{p}_t^1\|_2 - 2 \bar{r}_d, \|\mathrm{p}_t^e - \mathrm{p}_t^2\|_2 - 2\bar{r}_d, \\ 
    & \quad\quad\quad\quad\quad\quad\quad\quad\quad\quad \|[\mathrm{p}_t^e (1);\mathrm{p}_t^e (2)]-\bar{b}\|_2 - \bar{r}_d\bigg), \\
    &g(\mathrm{x}_t) = \bar{r} -\|\mathrm{p}_t^2-\bar{\mathrm{p}}\|_2,
\end{align*}

\normalsize\noindent
where $\bar{r}_d$ is the radius of the VTOL, $\bar{b}\in\mathbb{R}^2$ is the position of the building, $\bar{p}$ is the position of the target, and $\bar{r}$ is the radius of the target.

\subsubsection{Nonlinear Harvester Unloading (Harvester)}
\label{subsubsec:Harvester}
Consider a tractor unloading crops from a combine harvester in motion. The tractor must pull alongside the harvester and stay there while unloading the harvester without running over crops or hitting the harvester. At any moment, the harvester may speed up or slow down depending on terrain, field conditions, or crop density.

We model this nonlinear system as follows:
\begin{align}
    % \begin{split}
    &\mathrm{x}_{t+1}(1) = \mathrm{x}_t(1) + \Delta t(\cos(\mathrm{x}_t(3))\mathrm{x}_t(4) - \mathrm{x}_t(5) - v_c),\notag\\
    &\mathrm{x}_{t+1}(2) = \mathrm{x}_t(2) + \Delta t \sin(\mathrm{x}_t(3))\mathrm{x}_4(t),\notag\\
    &\mathrm{x}_{t+1}(3) = \mathrm{x}_t(3) + \Delta t u_t(1),\notag\\
    &\mathrm{x}_{t+1}(4) = \mathrm{x}_t(4) + \Delta t u_t(2),\notag\\
    &\mathrm{x}_{t+1}(5) = \mathrm{x}_t(5) + \Delta t d_t,
    % \end{split}
\end{align}
where $\mathrm{x}_t(1)$ is the horizontal distance from the tractor to the harvester, $\mathrm{x}_t(2)$ is the vertical distance from the tractor to the harvester, $\mathrm{x}_t(3)$ is the angle of the tractor, $\mathrm{x}_t(4)$ is the speed of the tractor, $\mathrm{x}_t(5)$ is the portion of the harvester's velocity set by the disturbance, and $v_c$ is a constant velocity of the harvester. The control $u_t(1)$ is the angular velocity of the tractor and $u_t(2)$ is the acceleration of the tractor. Disturbance $d_t$ is the acceleration of the harvester. $g$ is defined as:
\begin{equation}
    g(\mathrm{x}_t) = \bar{r}_1-\|[\mathrm{x}_t(1);\mathrm{x}_t(2)] -[0,\bar{o}_c]\|_2,
\end{equation}
where $\bar{o}_c$ is the location of the target at $-2$, and $\bar{r}_1$ is the target radius of 1.5.

Avoid includes the field of crops yet to be harvested, the harvester, and the crops in front of the harvester. It is characterized by:
$l_1(\mathrm{x}_t),l_2(\mathrm{x}_t),l_3(\mathrm{x}_t)<0$, where
\begin{equation}
    \begin{split}
        &l_1(\mathrm{x}_t) = -\mathrm{x}_t(2) - \bar{r}_2+2\bar{r}_3, \\
        &l_2(\mathrm{x}_t) = \min(|\mathrm{x}_t(1)-50|-50, |\mathrm{x}_t(2)|-(\bar{r}_3+\bar{r}_2)), \\
        &l_3(\mathrm{x}_t) = \|[\mathrm{x}_t(1);\mathrm{x}_t(2)]\|_2 - (\bar{r}_2+\bar{r}_3),
    \end{split}
\end{equation}
where $\bar{r}_2$ is the radius of the tractor and $\bar{r}_3$ is the radius of the harvester. These can be combined into a single constraint inequality: $l(\mathrm{x}_t)=\min\{l_1(\mathrm{x}_t),l_2(\mathrm{x}_t),l_3(\mathrm{x}_t)\}>0$.

\subsubsection{Two-Step Deep Reinforcement Learning Training}
\label{sec:twoStepDDPG}

When implementing our method with DDPG for the VTOL taxi and harvester examples, we found the following general principles helpful for achieving a well-trained policy:
\begin{itemize}
   \item Low Critic Loss: A well-trained policy typically exhibits a critic loss ranging from 0.01 to 0.001, with lower values indicating greater accuracy. Critic loss must converge before effective policy training can occur.
   \item Training Duration: We found it advantageous to continue training even after the loss and reward values appear stable. This additional training reduces the likelihood of outliers or edge cases during evaluation.
   \item Importance of training $H$: A well trained $H$ value function is critical for training $V$. Any training error from $H$ will be amplified in $V$. Generally, when the level-zero set of $H$ is completely inside the target set, it can be used for training $V$.
   \item Episode Lengths: $H$ trains best with episodes twice as long as those for $V$. At minimum, both $H$ and $V$ should have sufficiently long episodes to allow the agent to move from any state to the target set.
\end{itemize}
During offline training, we used one NVIDIA RTX A6000, and a AMD Ryzen Threadripper PRO 5975WX CPU, along with 256 GB of RAM. The training time of $H$ and $V$ depends on the batch and layer sizes, but were under three hours for both experiments. Exact values for the final tuned VTOL and Harvester models can be found in Table \ref{tab:hyperparam}. Once trained, the RAS framework is more than capable of running in real time. The final VTOL RAS policy is able to execute at 963.4 Hz, and the Harvester RAS policy executes at 1,366.1 Hz on the same hardware used for training. Hyperparameter settings for training the VTOL and Harvester environments are provided in Table \ref{tab:hyperparam}.

\begin{table}[]
\centering
\label{tab:hyperparam}
\begin{tabular}{@{}lll@{}}
\toprule
Hyperparameter                 & VTOL           & Harvester    \\ \midrule
Control Layers (Actor, Critic) & 512x4, 768x4   & 512x3, 512x4 \\
Batch Size (H, V)              & 512, 512       & 512, 256     \\
Epochs                         & 100            & 100          \\
Environment Steps Per Epoch    & 40,000         & 40,000       \\
Training Time (H\,/\,V) [min]  & 123.15\,/\,162.93 & 135.88\,/\,114.27 \\ \bottomrule
\end{tabular}
\caption{Training Hyperparameters}
\vspace{-2em}
\end{table}

\subsubsection{Justification of the Two Examples for Demonstrating the RAS task}
% Environment Justification, Environment Set Differences, Simulation System Set Differences, Source of Differences Between T and AS
The VTOL and Harvester environments have large differences between $T$ and $\mathcal{AS}$ enabling a clear demonstration of how RA policies fail the RAS task. In VTOL, the high maximum velocity compared to low maximum acceleration results in many states being unable to slow within target. In Harvester, the tractor must match the speed and direction of the Harvester, limiting the number of directions from which it can enter $T$ and stay.

\subsection{Claim Validation}
\label{subsec:baseline}

\begin{table}
    \centering
    \begin{tabular}{@{}|lll@{}}
        \toprule
        \multicolumn{1}{c|}{Method} & VTOL & Harvester \\ \midrule
        \multicolumn{1}{l|}{Our RAS Policy RAS Task Success Rate}& 93.0 & 99.3 \\ \midrule
        \multicolumn{1}{l|}{Baseline RA Policy RAS Task Success Rate} & 0 & 73.5 \\ \midrule
        \multicolumn{1}{l|}{Our Identified $\mathcal{RAS}$ Accuracy} & 95.8 & 99.4 \\ \bottomrule
    \end{tabular}
    \caption{Our RAS framework achieves higher success rates than the baseline RA method, and provides an accurate identified $\mathcal{RAS}$.}
\label{tab:cityResults}
\vspace{-3em}
\end{table}

\subsubsection{\textbf{Demonstration of Claim 2}}
To validate that our identified $\mathcal{RAS}$ was accurate, we used the policy and adversarial disturbances from our framework to generate 2,000 trajectories, each 500 time steps long. Half of the trajectories were sampled from within our framework's identified $\mathcal{RAS}$, and the others were from outside the identified $\mathcal{RAS}$. A trajectory was considered to have performed the RAS task if it safely entered $T$ and remained there without leaving even momentarily. We measure accuracy as the fraction of states for which our identified $\mathcal{RAS}$ correctly predicts the ability to safely reach and stay within the target set.

Our framework was highly successful at correctly identifying $\mathcal{RAS}$, the results of which can be found in the last row of Table \ref{tab:cityResults}. Our framework achieved 95.8\% accuracy in VTOL and 99.4\% accuracy in Harvester. 

\subsubsection{\textbf{Demonstration of Claim 3}}
To compare our RAS framework with a baseline RA method \cite{li_infinite-horizon_2024}, we sampled 1,000 initial states from within the identified $\mathcal{RAS}$. We then used each state as a starting point for the RA and RAS policies. We defined the success rate as the percentage of trajectories that successfully performed the RAS task. Our framework achieved a 93\% success rate in VTOL and 99.3\% success rate in Harvester compared to the baseline RA method's success rates of 0\% and 73.5\% in VTOL and Harvester, as shown in the second and third rows of Table~\ref{tab:cityResults}.\footnote{The RA policy achieved high success rates for safely reaching $T$ of 98.9\%, and 100\% across the VTOL, and harvester experiments.}

% Why RA fails
\subsection{Intuitive Explanation of RA Failure}
\label{subsec:RAFail}
\begin{figure}
    \centering
    \includegraphics[width=1\linewidth]{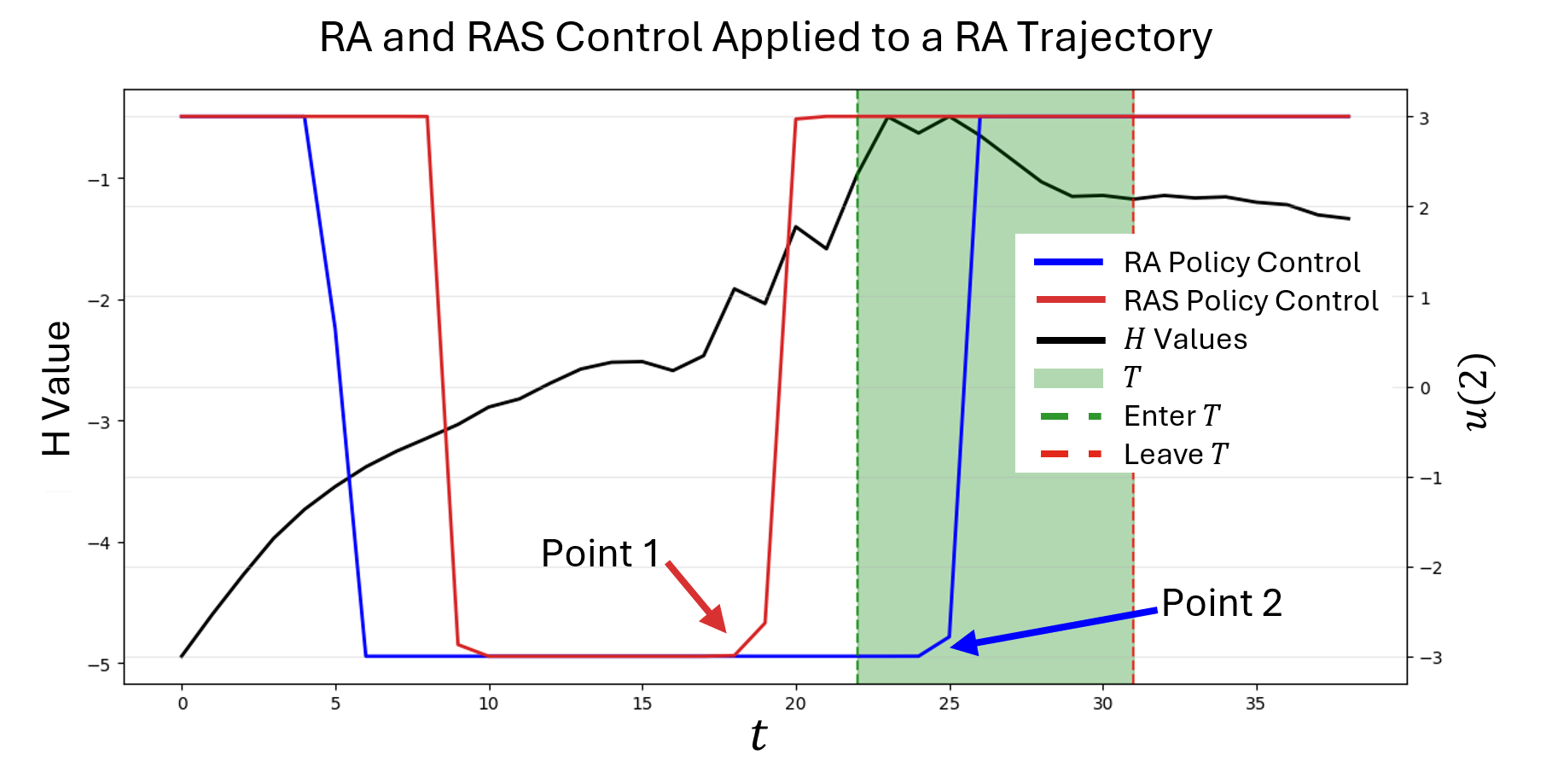}
    \caption{
    The blue and red lines represent the acceleration control outputs of the RA and RAS policies along the trajectory generated by the RA policy used in Fig. \ref{fig:combineTraj} for the Harvester example.%}
    The black line represents $H$ values across the trajectory. The $H$ values are negative, indicating that RA has not entered the $\mathcal{AS}$ and is therefore unable to stay. Additionally, the RAS policy indicates that the system should start to slow at point 1 when the tractor is outside $T$, but the RA policy waits until point 2 when the system is in the middle of $T$.}
    \label{fig:combineControl}
\end{figure}
\begin{figure}
    \centering
    \includegraphics[width=1\linewidth]{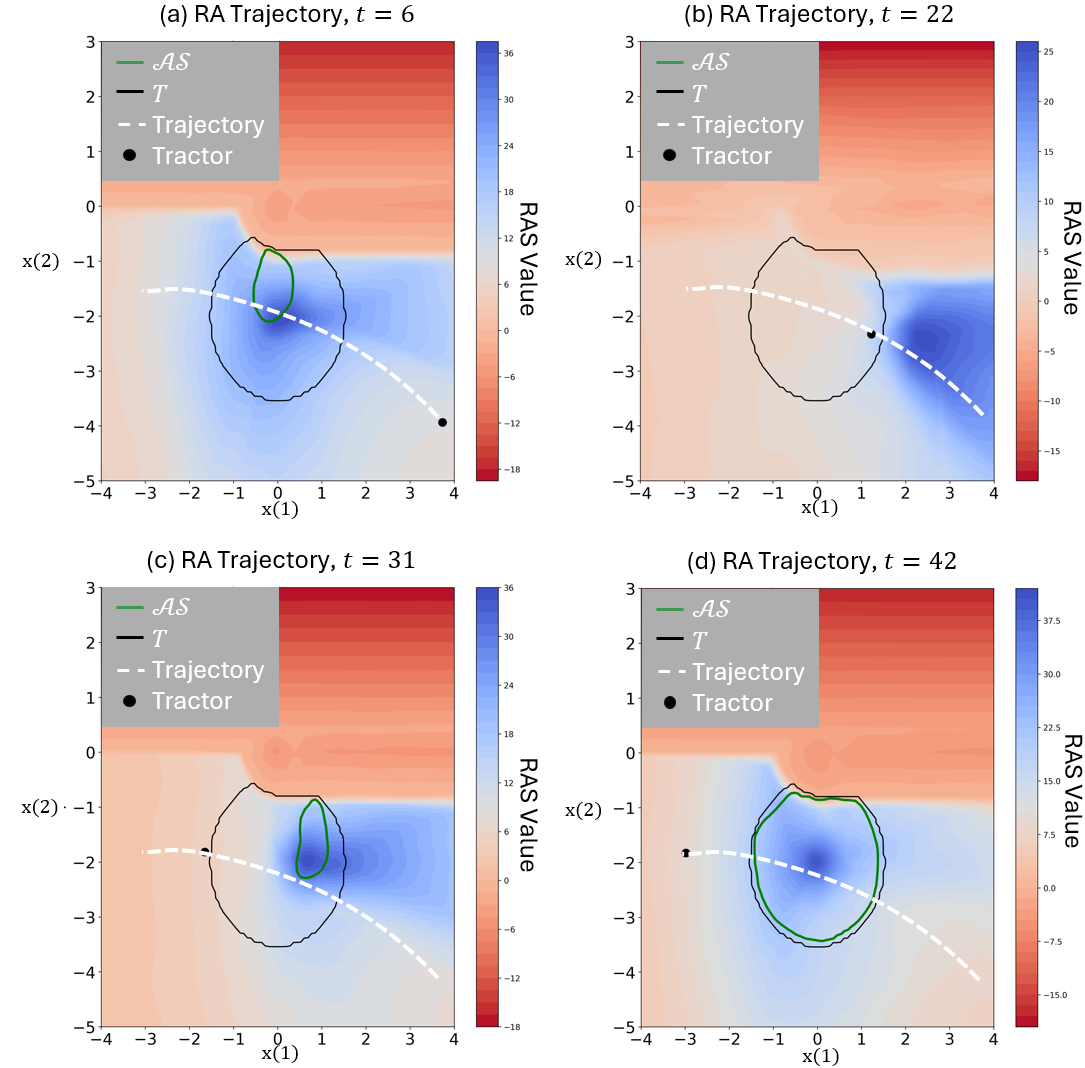}
    \caption{A Sequence of states for a Harvester example trajectory generated by the RA policy at  t = 9, 22, 31, and 42. The black outline represents $T$. The green outline represents the $\mathcal{AS}$. The trajectory never enters $\mathcal{AS}$, and exits $T$.}
    \label{fig:combineTraj}
    \vspace{-1em}
\end{figure}
The RA policy fails at the RAS task because it reaches $T$, but not $\mathcal{AS}$. An example of this is shown in Fig. \ref{fig:combineTraj}. Although the tractor does pass through $T$, the plot of $H$ values for this trajectory in Fig. \ref{fig:combineControl} never reaches zero, confirming that it never enters $\mathcal{AS}$. This is because the RA policy does not start to slow the tractor until it has reached Point 2 in Fig. \ref{fig:combineControl} as shown by the blue line representing acceleration control from the RA policy. 

To examine whether the system can slow earlier to reach $\mathcal{AS}$ and remain within $T$, we plot the acceleration control signal from the RAS policy along the same trajectory driven by the RA policy (shown in red in Fig. \ref{fig:combineControl}). We refer to this as the RAS policy control. This indicates that the tractor should begin to slow before entering $T$ at Point 1 to reach $\mathcal{AS}$.

The control difference between the RA and RAS policies is a result of the RA policy using $T$ as its target and the $\pi_V$ using $\mathcal{AS}$ as its target. $\mathcal{AS}$ has speed bounds which inform $\pi_V$ what speeds it must reach within $T$ for $\pi_H$ to be able to stay. However, $T$ has no such restrictions, leading the RA policy to reach $T$ as quickly as possible, at maximum speed. For static targets, unless the dynamics of a specific system allow it to reverse direction from maximum speed within half of $T$, the RA policy will miss $\mathcal{AS}$ and leave $T$.

% \begin{figure}
%     \centering
%     \includegraphics[width=0.7\linewidth]{figures/vtolTraj.png}
%     \caption{A: The blue and red lines represent control from the  RA policy  and RAS policy  applied to the RA trajectory in B.  The black line represents $H$ across the RA trajectory. Note that 1: The RA policy does not start slowing the system until it is in the middle of $T$ at point 2, but the RAS policy would have slowed at point 1 before it reached $T$. 2: The black line representing $H$ is is never greater than 0, so RA has not reached the $\mathcal{AS}$. B: Trajectories from the RA and RAS policies starting from the same initial state. The yellow line represents the path of the controlled ego VTOL. The green and red are the paths of the other VTOLs. The gray cylinder represents a static building. The RAS policy is able to successfully perform the RAS task and stay within the yellow target set, however the RA policy never reaches $\mathcal{AS}$ and is therefore unable to stay.}
%     \label{fig:vtolTraj}
% \end{figure}

% Set Differenc
\subsection{$\mathcal{RA}$ and $\mathcal{RAS}$ Set Differences:}
\begin{figure}
    \centering
    \includegraphics[width=1\linewidth]{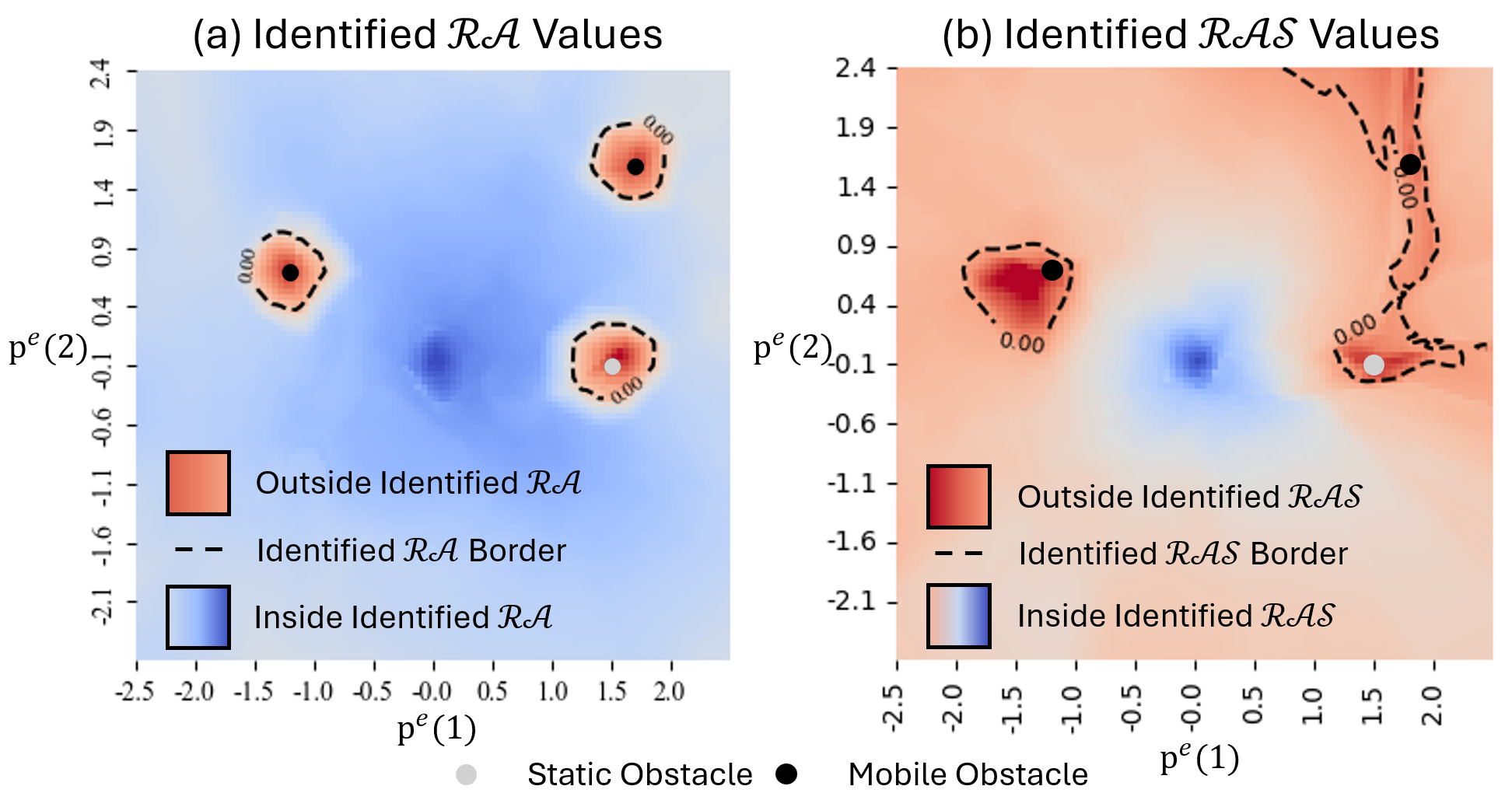}
    \caption{The identified $\mathcal{RAS}$ and $\mathcal{RA}$ for the VTOL example. The black circles represent the mobile obstacles, and the gray circle represents the static obstacle. The identified $\mathcal{RA}$ in (a) is a super-set of the identified $\mathcal{RAS}$ in (b) due to the limits the RAS policy places on the system to enable it to stay.}
    \label{fig:vtolSet}
    \vspace{-1em}
\end{figure}
% \begin{figure}
%     \centering
%     \includegraphics[width=1\linewidth]{figures/combineSets.png}
%     \caption{The constant states for the tractor are $\mathrm{x}(3)=90$ (pointed directly in positive $\mathrm{x}(2)$), $\mathrm{x}(4)=3$. This results in areas near the unharvested crops being outside of both the identified $\mathcal{RA}$ in (a) and $\mathcal{RAS}$ in (b). Note that while $T$ is completely with in the identified $\mathcal{AS}$, the top section of $T$ is outside of $\mathcal{RAS}$.}
%     \label{fig:combineSet}
% \end{figure}
%In theory, with identical target and constraint sets, the RA set will be a super set of the RAS set. In practice, even with learning errors, our results largely confirm this. One example for the Harvester appears in Fig. \ref{fig:combineSet} (a) and (b). The RA task does not consider any time steps after it reaches $T$, and therefore considers any state within $T$ to also be within the identified $\mathcal{RAS}$ in Fig. \ref{fig:combineSet} (a). However, the RAS task requires the system to stay within $C$ for all time. Therefore, even states within $T$ can be outside of the identified $\mathcal{RAS}$ set as shown in \ref{fig:combineSet} if they could result in leaving $T$ or $C$.
In theory, with identical target and constraint sets, the RA set will be a super set of the RAS set. In practice, even with learning errors, our results largely confirm this. The identified $\mathcal{RA}$ and $\mathcal{RAS}$ for the VTOL environment can be seen in Fig. \ref{fig:vtolSet}. It can be observed that the RAS policy is generally more cautious around both the static and mobile obstacles.

\addtolength{\textheight}{-0cm}   % This command serves to balance the column lengths

\section{Conclusion} 
\label{sec:conclusion}
We propose a two-step deep reinforcement learning method to solve the Reach-Avoid-Stay (RAS) problem. Our framework addresses the limitations of multiple prior methods including the difficulty of designing a CLBF \cite{meng_lyapunov-barrier_2022}, and the funnel method \cite{das_funnel-based_2023} assumption that the system can instantly stop. We demonstrate that without training error, our approach yields the maximal robust RAS set, which is a super-set of prior method's RAS sets. Even when facing training error, we identify the maximal robust RAS set with over 95\% accuracy. We demonstrate a higher success rate than a reach-avoid baseline \cite{li_certifiable_2024} in the RAS task, and explain how fundamental differences between the RA policy and RAS policy leads to the RA policy being unable to perform the RAS task.

%The key limitation of our framework its inability to handle unknown environments. For example, $\mathcal{RAS}$ and $\pi_\mathcal{RAS}$ can be computed for a warehouse, but must be completely recomputed if the warehouse layout changes. Modern robots use real-world sensor data to enable their control algorithms to be applied to any environment the robot encounters. %\cite{macenski_marathon_2020} IMO we could cite NAV2 here if we wanted to given that it is the prime example of doing this, but it is not necisary. Future work will focus on integrating this real-world sensor data with our framework to enable efficient online computation of $\mathcal{RAS}$ and $\pi_\mathcal{RAS}$.

The primary limitation of our framework lies in its reliance on prior knowledge of the environment, including target sets, constraint sets, and system dynamics. When this information does not accurately reflect real-world conditions, the trained policies cannot guarantee successful completion of the RAS task. This challenge has recently been addressed in safety applications by approaches such as reinforcement learning that introduces additional parameters to capture unknown environment information \cite{lin2024one}, or by integrating safety analysis with online control methods \cite{borquez2025dualguard}. However, these approaches still lack rigorous theoretical analysis of safety and performance assurances. In future work, we aim to close this gap by extending reachability-based RL and incorporating real-world sensor data into our framework.

%%%%%%%%%%%%%%%%%%%%%%%%%%%%%%%%%%%%%%%%%%%%%%%%%%%%%%%%%%%%%%%%%%%%%%%%%%%%%%%%
% \input{Appendix.tex}
%%%%%%%%%%%%%%%%%%%%%%%%%%%%%%%%%%%%%%%%%%%%%%%%%%%%%%%%%%%%%%%%%%%%%%%%%%%%%%%%

\bibliographystyle{IEEEtran}
\bstctlcite{references}
\bibliography{references}

\end{document}